\documentstyle[12pt]{article}

\newlength{\dinwidth}
\newlength{\dinmargin}
\def\lapproxeq{\lower .7ex\hbox{$\;\stackrel{\textstyle
<}{\sim}\;$}}
\def\gapproxeq{\lower .7ex\hbox{$\;\stackrel{\textstyle
>}{\sim}\;$}}

\begin{document}
\titlepage
\begin{flushright}
hep-ph/9503266  \\
DTP/95/22  \\
March 1995 \\
\end{flushright}

\begin{center}
\vspace*{2cm}
{\Large {\bf The gluon distribution at small $x$ obtained from a
unified evolution equation}} \\
\vspace*{1cm}
J.\ Kwieci\'{n}ski\footnote{On leave from Henryk
Niewodnicza\'{n}ski Institute of Nuclear Physics, 31-342
Krak\'{o}w, Poland.} and A.\ D.\ Martin,

Department of Physics, University of Durham, Durham, DH1 3LE,
England \\

and \\

P.\ J.\ Sutton,

Department of Physics, University of Manchester, Manchester, M13
9PL, England
\end{center}

\vspace*{3cm}
\begin{abstract}
We solve a unified integral equation to obtain the $x, Q_T$ and
$Q$
dependence of the gluon distribution of a proton in the small $x$
regime; where $x$ and $Q_T$ are the longitudinal momentum
fraction and the transverse momentum of the gluon probed at a
scale $Q$.  The equation generates a gluon with a steep $x^{-
\lambda}$ behaviour, with $\lambda \sim 0.5$, and a $Q_T$
distribution which broadens as $x$ decreases.  We compare our
solutions with, on the one hand, those that we obtain using the
double-leading-logarithm approximation to Altarelli-Parisi
evolution and, on the other hand, to those that we determine from
the BFKL equation.
\end{abstract}

\newpage

\noindent {\large \bf 1.  Introduction}

Understanding the details of the small $x$ behaviour of parton
distributions is one of the most challenging problems of
perturbative QCD \cite{glr,smx}.  Moreover this topic has
recently become of particular phenomenological interest with the
advent of measurements of deep inelastic scattering at the high
energy electron-proton collider, HERA, which have opened up the
small $x$ regime \cite{felt}.  There now exist data for the
proton structure function $F_2$ for $x$ as low as $x \sim
10^{-4}$.  These measurements reveal a significant rise of $F_2$
with decreasing $x$, which has been taken to signal novel
phenomena, although whether, in fact, this is the case or whether
conventional explanations suffice, remains to be settled.

First we recall the conventional treatment of \lq\lq hard"
scattering processes involving hadrons at moderate values of $x$,
say $x \gapproxeq 0.05$.  Then the observable quantities are
calculated in perturbative QCD using the mass factorization
theorem \cite{mft} in which the collinear singularities, which
occur in the partonic subprocesses, are absorbed into universal
parton densities.  To be specific let us take as an example the
longitudinal structure function\footnote{We choose $F_L$ as our
example, rather than $F_2$, because we are interested in hard
scattering observables which are driven directly by the gluon.
We could have used $F_2$ but then we would have to deal with the
collinear singularities of $\hat{F}_2$ associated with the $g
\rightarrow q\overline{q}$ transition.  These singularities are
absorbed into the universal sea quark distributions.  The
simplification in which we neglect the sea quarks would therefore
have been incomplete, even at leading order, and we would have
had to broaden the discussion.} of the proton $F_L (x, Q^2)$ and,
for simplicity, consider only the gluon\footnote{The gluon
dominates the other partons in the small $x$ regime, which is the
main concern of our study.} partonic constituent.  Then we have
\begin{equation}
F_L (x, Q^2) = \int_x^1 \frac{dx^\prime}{x^\prime}
g(x^\prime,Q^2) \hat{F}_L (x/x^\prime,\alpha_S(Q^2))
\label{b2}
\end{equation}
\noindent where we have chosen the mass factorization scale to be
the \lq\lq hard" scale $Q^2$ of the process.  The absorption of
the collinear logarithmic singularities make the gluon density
$g(x, Q^2)$ \lq\lq run", with a $Q^2$ dependence given by the
Altarelli-Parisi evolution equations \cite{ap}.  That is
perturbative QCD does not determine the gluon absolutely but only
its evolution from a non-perturbative input form.
Altarelli-Parisi evolution resums the leading $\alpha_S \log
(Q^2/Q_0^2)$ contributions.  In a physical gauge the $\alpha_S^n
\log^n (Q^2/Q_0^2)$ contribution can be associated with a
space-like chain of $n$ gluon emissions in which the successive
gluon transverse momenta are strongly ordered along the chain
\cite{dok}, that is $k_{T1}^2 \ll \ldots \ll k_{Tn}^2 \ll Q^2$.
The next-to-leading order contribution corresponds to the case
when a pair of gluons are emitted without strong $k_T$-ordering
(and iterations of this configuration).  Then we have a power of
$\alpha_S$ unaccompanied by $\log (Q^2/Q_0^2)$.  A key feature of
this conventional partonic approach is that the gluonic structure
function, $\hat{F}_L$ (which at lowest order arises from the
subprocess $\gamma^* g \rightarrow q\overline{q}$) is calculated
assuming that the incoming gluon has negligible transverse
momentum (and hence virtuality) as compared to the scale of the
hard process.  That is, on account of strong-ordering, we are
able to work in terms of the density $g(x,Q^2)$ of the gluon
integrated over its transverse momentum $Q_T$.

At sufficiently high electron-proton c.m. energy, $\sqrt{s}$, we
encounter a second large variable, $1/x \sim s/Q^2$, and in this
small $x$ regime we must resum the leading $\alpha_S \log
(1/x)$ contributions.  The key ingredients of the QCD framework
in this regime are the high-energy $Q_T$-factorization theorem
\cite{ktf} and the BFKL equation \cite{bfkl,jar} for the gluon
distribution, $F(x,Q_T)$, unintegrated over its transverse
momentum $Q_T$.  Using $Q_T$-factorization our sample observable
is now given by
\begin{equation}
F_L(x, Q^2) = \int_x^1 \frac{dx^\prime}{x^\prime} \int \frac{d^2
Q_T}{\pi} F(x^\prime,Q_T) \tilde{F}_L \biggl (\frac{x}{x^\prime},
\frac{Q_T^2}{Q^2}, \alpha_S(Q^2) \biggr )
\label{c2}
\end{equation}
\noindent where $\tilde{F}_L$ denotes the gluonic structure
function calculated using an off-shell ($Q_T^2 \neq 0$) gluon.
If we were to return to the strongly-ordered transverse momentum
configuration then
\begin{equation}
\tilde{F}_L \rightarrow \hat{F}_L = \tilde{F}_L (x/x^\prime, 0,
\alpha_S
(Q^2))
\label{d2}
\end{equation}
\noindent and
\begin{equation}
xg(x,Q^2) = \int \frac{d^2 Q_T}{\pi} F(x, Q_T) \Theta (Q - Q_T),
\label{a2}
\end{equation}
though, of course, it is inappropriate to work in terms of
the familiar integrated distribution $g(x,Q^2)$ at small $x$.

The BFKL equation for the unintegrated gluon density $F(x,Q_T)$
sums the leading $\alpha_S \log (1/x)$ contributions.  The
strong-ordering in transverse momenta is no longer applicable and
we now have a \lq\lq random walk" or diffusion in $k_T$ as we
proceed along the chain.  The enlarged $k_T$ phase space leads to
a $x^{-\lambda}$ growth, with decreasing $x$, where $\lambda \sim
0.5$.  Indeed the observed behaviour of $F_2$ appears consistent
with the precocious onset of this leading $\log (1/x)$ behaviour
of the gluon \cite{akms}, but the definitive confirmation that
this is the case must await the calculation of the subleading
corrections to the BFKL equation.  Clearly the BFKL equation,
which resums the leading $\log (1/x)$ contributions, has a
limited region of validity.  In principle, it is restricted to
the region $\alpha_S \log (1/x) \sim {\cal O}(1)$ and $\alpha_S
\log (Q^2/Q_0^2) \ll 1$, where $Q_0^2$ indicates the boundary of
the non-perturbative domain, $Q_0^2 \sim 1$ GeV$^2$.  To make
progress we need to know how the BFKL formalism links with the
conventional Altarelli-Parisi dynamics at larger $x$ and large
$\log (Q^2/Q_0^2)$.

A theoretical framework which gives a unified treatment
throughout the $x, Q^2$ kinematic region has been provided by
Catani, Ciafaloni, Fiorani and Marchesini [11-15].  The resulting
equation, which we shall call the CCFM equation, treats both the
small and large $x$ regions in a unified way.  The equation is
based on the coherent radiation of gluons, which leads to an
angular ordering of the gluon emissions along the chain.  In the
leading $\log (1/x)$ approximation the CCFM equation reduces to
the BFKL equation, whereas at moderate $x$ the angular ordering
becomes an ordering in the gluon transverse momenta and the CCFM
equation becomes equivalent to standard Altarelli-Parisi
evolution \cite{ap}.  The angular ordering introduces an
additional scale (which turns out to be essentially the hard
scale $Q$ of the probe), which is needed to specify the maximum
angle of gluon emission.  Thus we must work with a
scale-dependent, unintegrated gluon density $F(x, Q_T, Q)$.  At
very small $x$ the angular ordering does not provide any
constraint on the transverse momenta along the chain and $F$
becomes the $Q$-independent gluon of the BFKL equation.

The aim of our paper is to study the CCFM equation in detail and,
in particular, to obtain numerical solutions so as to reveal the
small $x$ behaviour of the gluon distribution in the proton.  In
Section 2 we sketch the derivation of the CCFM equation, while in
Section 3 we study its properties in the important small $x$
region.  We make approximations which both simplify the
discussion and facilitate the solution of the equation.  To gain
insight we study both the \lq\lq folded" and \lq\lq unfolded"
versions of the equation.  In Section 4 we present numerical
solutions $F(x, Q_T, Q)$ of the CCFM equation and compare them
with solutions that we obtain from solving the two limiting
versions of the equation.  That is we compare the CCFM solutions
with those obtained from (i) the double-leading-logarithm
approximation to the Altarelli-Parisi equation and (ii) the BFKL
equation.  Section 5 contains our conclusions on the small $x$
behaviour of the gluon.

\vspace*{4mm}
\noindent {\large \bf 2.  Coherent branching : the master
equation for the gluon}

The perturbative evaluation of physical QCD quantities in
general, and parton distributions in particular, is complicated
by the presence of large logarithms which arise from the emission
of both soft and collinear gluons.  The origin of the large
logarithms can be seen from Fig. 1.  The differential probability
for emitting a gluon of 4-momentum $q$ is of the form
\begin{equation}
dP \sim \alpha_S \frac{dz_g}{z_g} \frac{dq_T^2}{q_T^2}
\label{a1}
\end{equation}
\noindent where $q_T$ is the transverse momentum and $z_g$ is the
longitudinal momentum expressed as a fraction of the momentum of
the parent gluon.

Here we focus attention on the gluon distribution within a
proton.  To predict the correct behaviour of the distribution it
is necessary to resum the large logarithms which arise not just
from single but from multigluon emissions to all orders in
$\alpha_S$.  A typical contribution is shown in Fig. 2, where a
gluon of low space-like virtuality evolves to higher virtuality
and lower energy by successive gluon emission.  It can be shown
that the emissions are coherent in the sense that there is
angular ordering, $\theta_i > \theta_{i-1}$, along the chain,
where $\theta_i$ is the angle that the $i^{th}$ gluon makes to
the original direction [11-16].  Outside this region there is
destructive interference such that the multigluon contributions
vanish to leading order.  We speak of coherent branching.

As mentioned in Section 1, due to the presence of angular (rather
than transverse momentum) ordering of the emitted gluons we need
to expose the transverse momentum, $Q_T$, dependence of the
probed gluon.  That is we work in terms of the scale ($Q$)
dependent \lq\lq unintegrated" gluon density $F(x, Q_T, Q$),
which specifies the chance of finding a gluon with longitudinal
momentum fraction $x$ and transverse momentum of magnitude $Q_T$
\cite{march1}.  The integral equation for $F(x, Q_T, Q)$, which
effects the summation of the large logarithms, can be
approximated on the one hand, to yield the BFKL equation at small
$x$, where $F$ becomes independent of $Q$ and, on the other hand,
to yield at moderate $x$ the Altarelli-Parisi (or GLAP) evolution
equation for the integrated distribution $g(x,Q^2)$.  It is this
master equation which we wish to investigate and to solve.

It is necessary to outline the derivation of the equation
\cite{ciaf,cat1,march1}.  A crucial development has been the
proof of the soft gluon factorization theorems which allow the
inclusion, not only real gluon emission, but also the virtual
emission contributions which tame the singular behaviour in the
two boundary regions $z \rightarrow 0$ and $z \rightarrow 1$.
A recurrence relation can then be obtained which expresses
the contribution from $n$-gluon emission in terms of that from $n
- 1$ emission.  In essence, we integrate the differential
probability $dP$ for the emission of the extra gluon over the
relevant region of phase space, where (\ref{a1}) takes the
explicit form
\begin{equation}
dP = \Delta_S \tilde{P} dz \frac{dq_T^2}{q_T^2} \Theta (\theta -
\theta^\prime).
\label{a3}
\end{equation}
Here $\Theta(\theta - \theta^\prime)$ reflects the angular
ordering and $\tilde{P}$ is the gluon-gluon splitting function
\begin{equation}
\tilde{P} = {\overline \alpha}_S \left [ \frac{1}{1 - z} +
\Delta_{NS} \frac{1}{z} - 2 + z(1 - z) \right ].
\label{a4}
\end{equation}

\noindent We define ${\overline \alpha}_S = C_A \alpha_S/\pi =
3\alpha_S/\pi$.  The multiplicative factors $\Delta_S$ and
$\Delta_{NS}$, known as the Sudakov and non-Sudakov form factors,
arise from the resummation of the virtual corrections.  They
cancel the singularities manifest as $z \rightarrow 1$ and $z
\rightarrow 0$ respectively.  These form factors have exponential
forms which we present in eqs. (\ref{a7}) and (\ref{a8}) below.

We use (\ref{a3}) to obtain a recursion relation expressing the
distribution ${\cal{F}}_n$ in terms of ${\cal{F}}_{n-1}$.  The
distribution ${\cal F}_n (x, Q_T, z, q)$ corresponds to the
$n$-rung ladder diagram, where the variables are defined in Fig.\
3.  Following refs.\ \cite{march1,brw1} we impose the
angular-ordered
constraint by introducing rescaled transverse momenta
\begin{equation}
q \equiv \frac{q_T}{1 - z} \approx \theta E^\prime, \quad
q^\prime \equiv \frac{q_T^\prime}{1 - z^\prime} \approx
\theta^\prime E^{\prime\prime}
\label{a5}
\end{equation}
\noindent where $1 - z$ is the longitudinal momentum fraction of
the gluon emitted at angle $\theta$ and $E^\prime$ is the energy
component of the exchanged gluon with spacelike momentum
$x^\prime
p$.  Here we have used the small angle approximation, $\tan\theta
\approx \theta$.  The coherence constraint $\theta >
\theta^\prime$ therefore implies $q > z^{\prime}q{^\prime}$ and
so (\ref{a3}) gives
\begin{equation}
{\cal{F}}_n (x, Q_T, z, q) = \int _{x/z}^1 dz^\prime \int
\frac{d^2q^\prime}{\pi q^{\prime 2}} \Theta (q- z^\prime
q^\prime) \Delta_S (q, z^\prime q^\prime) \tilde{P}(z, q, Q_T)
{\cal{F}}_{n-1} \left ( \frac{x}{z}, Q_T^\prime, z^\prime,
q^\prime
\right ),
\label{a6}
\end{equation}
\noindent see Fig. 3.  Due to the presence of angular ordering,
it appears that we have to consider a less inclusive structure
function than the unintegrated distribution $F(x, Q_T, Q)$
itself.  To be precise we have exposed not only the $x, Q_T$ of
the probed gluon, but also the $z, q$ dependence which
specifies the previous emitted gluon.  Note that the variable
$Q_T^\prime$ in ${\cal{F}}_{n-1}$ is the magnitude of the vector
sum $\mbox{\boldmath $Q$}_T + (1 - z) \mbox{\boldmath $q$}$ and,
so, in principle, the angular integration in $d^2q^\prime$ is
non-trivial.

The Sudakov form factor is given by
\begin{equation}
\Delta_S (q,z^\prime q^\prime) = \exp \left ( - \int_{(z^\prime
q^\prime)^2}^{q^2} \frac{dk^2}{k^2} \int_0^1 dx \frac{{\overline
\alpha}_S}{1 - z} \right ).
\label{a7}
\end{equation}
\noindent The region of integration corresponds to the
angular-ordered region from the angle $\theta^\prime$ of emission
of the $(n - 1)^{th}$ gluon to $\theta$ of the $n^{th}$ gluon.
Indeed $\Delta_S$ can be interpreted as the probability for not
emitting a gluon in this angular region.  This observation is
consistent with (\ref{a3}) which gives the differential
probability for {\bf single} gluon emission in an element of
phase space specified by $z, q_T^2$; the factor $\Delta_S$
ensuring that there is no prior emission.

The splitting function $\tilde{P}(z,q,Q_T)$ is given by
(\ref{a4})
in which the $1/z$ singularity is screened by virtual corrections
contained in the non-Sudakov form factor
\begin{eqnarray}
\Delta_{NS}(z,q,Q_T) & = & \exp \left ( - {\overline \alpha}_S
\int_z^{z_0} \frac{dz^\prime}{z^\prime} \int \frac{dk^2}{k^2}
\Theta
(Q_T^2 - k^2) \Theta (k - z^\prime q) \right ) \label{a8} \\
& = & \exp \left ( - {\overline \alpha}_S \log \left (
\frac{z_0}{z} \right ) \log \left ( \frac{Q_T^2}{z_0zq^2} \right
) \right ),
\label{b8}
\end{eqnarray}

\noindent where
$$
z_0 = \left \{ \begin{array}{lll}
1 & {\rm if} & (Q_T/q) \geq 1 \\
Q_T/q & {\rm if} & z < (Q_T/q) < 1 \\
z & {\rm if} & (Q_T/q) \leq z.
\end{array} \right.
$$
\noindent Unlike $\Delta_S$, the non-Sudakov form factor
$\Delta_{NS}$ is not just a function of the branching variables,
but depends on the history of the cascade via
\begin{equation}
Q_T = |\mbox{\boldmath $q$}_T + \mbox{\boldmath $q$}_T^\prime +
\mbox{\boldmath $q$}_T^{\prime\prime} +
\ldots |.
\label{a9}
\end{equation}
\indent Actually the recursion relation (\ref{a6}) is satisfied
by a more inclusive distribution $F_n(x, Q_T, Q)$ in which the
$z,q$ dependence of ${\cal{F}}_n$ is integrated over, subject to
the maximum angle specified by $Q$ \cite{march1}.  That is, if we
introduce a
distribution for $n$-gluon emission defined by
\begin{equation}
F_n(x, Q_T, Q) \equiv \int_x^1 dz \int \frac{d^2 q}{\pi q^2}
\Theta (Q - zq) \Delta_S (Q, zq) {\cal{F}}_n (x, Q_T, z,
q),
\label{a10}
\end{equation}
\noindent then the recursion relation (\ref{a6}) becomes
\begin{equation}
F_n(x, Q_T, Q) = \int_x^1 \frac{dz}{z} \int \frac{d^2
q}{\pi q^2} \Theta (Q - zq) \Delta_S (Q, zq) \tilde{P} (z, q,
Q_T) F_{n-1} \left ( \frac{x}{z}, Q_T^\prime, q \right ).
\label{a11}
\end{equation}
\noindent Finally we obtain the (scale-dependent unintegrated)
gluon density by summing over all gluon emissions
\begin{equation}
F(x, Q_T, Q) = \sum_{n = 0}^\infty F_n (x, Q_T, Q).
\label{a12}
\end{equation}
\noindent From (\ref{a11}) we find
\begin{eqnarray}
F(x, Q_T, Q) & = & F^0 (x, Q_T, Q) + \nonumber \\
& & + \int_x^1 dz \int \frac{d^2 q}{\pi q^2} \Theta (Q - zq)
\Delta_S (Q, zq) \tilde{P} (z, q, Q_T) F \biggl (\frac{x}{z},
Q_T^\prime, q \biggr ),
\label{a13}
\end{eqnarray}
\noindent with $Q_T^\prime = |\mbox{\boldmath $Q$}_T + (1 - z)
\mbox{\boldmath $q$}|$.  The inhomogeneous or \lq\lq no-rung"
contribution, $F^0$, may be regarded as the non-perturbative
driving term.  This basic integral equation (\ref{a13}), for the
gluon structure function $F (x, Q_T, Q)$, which we have called
the CCFM equation after its originators, is the starting point of
our analysis.  It may be approximated both at moderate $x$ to
yield the Altarelli-Parisi evolution equation, and at small $x$
to yield the BFKL equation, as we shall now show.

\vspace*{4mm}
\noindent {\large \bf 3.  Approximations in the small $x$ region}

To explore the structure of the gluon in the small $x$ region we
may approximate (\ref{a13}) by
\begin{equation}
F (x, Q_T, Q) = F^0 (x, Q_T, Q) + {\overline \alpha}_S \int_x^1
\frac{dz}{z} \int \frac{d^2 q}{\pi q^2} \Theta (Q - zq)
\Delta_{NS} (z, q, Q_T) F \left ( \frac{x}{z}, |\mbox{\boldmath
$Q$}_T + \mbox{\boldmath $q$}|, q \right )
\label{a14}
\end{equation}
\noindent where we have set $\Delta_S = 1$ and retained only the
$1/z$ term in the splitting function $\tilde{P}$.  We have also
approximated $(1 - z)q$ by $q$ in the argument $Q_T^\prime$ of
$F$.  In this small $x$ limit, the variable $q$ reduces to the
transverse momentum $q_T$ of the emitted gluon, see eq.
(\ref{a5}).  This is the CCFM integral equation
\cite{ciaf,cat1,march1} which we solve to find the $x, Q_T$ and
$Q$ dependence of the gluon distribution.  The procedure that we
adopt and the results that we obtain are presented in Section 4.

Also in Section 4 we compare the results obtained from the CCFM
equation with those obtained in the double leading logarithm
(DLL) approximation in which (\ref{a14}) reduces to
\begin{equation}
F(x, Q_T, Q) = F^0 (x, Q_T, Q) + \overline{\alpha}_S
\int_x^1 \frac{dz}{z} \int \frac{d^2 q}{\pi q^2} \Theta (Q - q)
F \left ( \frac{x}{z}, |\mbox{\boldmath $Q$}_T + \mbox{\boldmath
$q$}|, q \right ).
\label{b14}
\end{equation}

\noindent To be precise the DLL approximation is obtained by
setting $\Delta_{NS} = 1$ and by replacing the angular ordering,
$\Theta (Q - zq)$, by ordering in transverse momentum, $\Theta (Q
- q)$.  This procedure becomes equivalent to the conventional DLL
approximation for $xg(x, Q^2)$ after we integrate over $Q_T$, as
in (\ref{a2}).  Here we have the advantage that we can also
display the $Q_T$ dependence of the gluon distribution.

Before we present our numerical results it is informative to gain
insight into the structure of the CCFM equation, (\ref{a14}).  We
can simplify the discussion by rewriting (\ref{a14}) in terms of
the moment function
\begin{equation}
F_\omega (Q_T, Q) = \int_0^1 dx \; x^{\omega - 1} F(x, Q_T, Q).
\label{a15}
\end{equation}
\noindent We obtain
\begin{equation}
F_\omega (Q_T, Q) = F_\omega^0 (Q_T, Q) + {\overline \alpha}_S
\int \frac{d^2 q}{\pi q^2} H_\omega (Q, Q_T, q) F_\omega
(|\mbox{\boldmath $Q$}_T + \mbox{\boldmath $q$}|, q)
\label{a16}
\end{equation}
\noindent where
\begin{eqnarray}
H_\omega (Q, Q_T, q) & = & \int_0^1 dz \; z^{\omega - 1} \Theta
(Q - zq) \Delta_{NS} (z, q, Q_T)  \\
\label{b16}
& = & \Theta (Q-q) \int_0^1 dz\,z^{\omega-1} \Delta_{NS}
(z,q,Q_T) \nonumber \\
& & + \; \Theta (q-Q) \int_0^{Q/q} dz\,z^{\omega-1} \Delta_{NS}
(z,q,Q_T).
\label{a17}
\end{eqnarray}
\noindent The curtailment of the range of integration in the
second term is seen to be a
direct consequence of angular ordering.

\vspace*{3mm}
\noindent {\bf (a)  Unfolding the equation}

It is useful to unfold the kernel of (\ref{a16}) so that the real
emission and virtual corrections terms appear on equal footing,
i.e. to the same order in $\alpha_S$.  On the one hand this will
allow us to make correspondence with the BFKL equation, and on
the other hand it will guide us to the correct structure of the
driving term $F^0 (x, Q_T, Q)$ to be used as input for the CCFM
equation, (\ref{a14}).  To unfold the kernel we first integrate
(\ref{a17}) by parts
\begin{eqnarray}
H_\omega(Q,Q_T,q) & = & \frac{1}{\omega}  \biggr[ \Theta(Q - q)
+ \left (Q/q \right )^\omega \Theta (q - Q) \Delta_{NS}
(z = Q/q, q, Q_T) \biggr] \nonumber \\
& & - \; \frac{\overline{\alpha}_S}{\omega}
\int_0^1 dz \; z^{\omega - 1} \int \frac{dk^2}{k^2} \Theta (Q -
zq) \Theta (k - zq) \Theta (Q_T^2 - k^2) \Delta_{NS} (z, q, Q_T)
\nonumber \\
& = & \frac{1}{\omega} \biggr[ \Theta (Q - q) + \left (Q/q \right
)^\omega \Theta (q - Q) \Delta_{NS} \left (z = Q/q, q, Q_T \right
) \biggr] \nonumber \\
& & - \; \frac{\overline{\alpha}_S}{\omega} \int
\frac{dk^2}{k^2} \Theta (Q_T^2 - k^2) H_\omega (\min
\{Q,k\}, q, Q_T)
\label{a18}
\end{eqnarray}
\noindent where the last term has been simplified using
(\ref{a17}).  We insert (\ref{a18}) in (\ref{a16}) and express
the final term in terms of $F_\omega - F_\omega^0$.  We obtain

\begin{eqnarray}
F_\omega(Q_T,Q) & = & F_\omega^0(Q_T, Q) +
\frac{\overline{\alpha}_S}{\omega} \int \frac{dq^2}{q^2} \Theta
(Q_T^2 - q^2) F_\omega^0 (Q_T, \min \{Q,q\}) \nonumber \\
& + & \frac{\overline{\alpha}_S}{\omega} \int \frac{d^2 q}{\pi
q^2} \biggr\{ \biggl[ \Theta (Q-q) + (Q/q)^\omega \Theta
(q-Q) \Delta_{NS} (z = Q/q, q, Q_T) \biggr] F_\omega
(|\mbox{\boldmath $Q$}_T + \mbox{\boldmath $q$}|, q) \nonumber \\
& - & \Theta (Q_T^2 - q^2) F_\omega (Q_T, \min \{Q,q
\}) \biggr\}
\label{a19}
\end{eqnarray}

\noindent It remains to unfold the non-Sudakov form factor
$\Delta_{NS}$.  In the Appendix A, we show that (\ref{a19}) then
becomes
\begin{eqnarray}
F_\omega(Q_T,Q) & = & \frac{1}{\omega} \hat{F}_\omega^0(Q_T, Q) +
\frac{\overline{\alpha}_S}{\omega} \int \frac{d^2 q}{\pi q^2}
\left [\Theta (Q - q) + \left (\frac{Q}{q} \right )^\omega \Theta
(q - Q) \right ] F_\omega (|\mbox{\boldmath $Q$}_T +
\mbox{\boldmath $q$}|, q) \nonumber \\
& & - \; \frac{\overline{\alpha}_S}{\omega} \int \frac{d^2 q}{\pi
q^2} \Theta (Q_T^2 - q^2) F_\omega (Q_T, q) \nonumber \\
& & + \; \frac{\overline{\alpha}_S}{\omega} \Theta (Q_T^2 - Q^2)
\int_{Q^2}^{Q_T^2} \frac{d^2 q}{\pi q^2} q^2 \frac{\partial
F_\omega (Q_T,q)}{\partial q^2} \log \left (\frac{q^2}{Q_T^2}
\right ) \left [ \left ( \frac{Q}{q} \right )^\omega - 1 \right ]
\label{a20}
\end{eqnarray}
\noindent where the driving term of the unfolded equation is
related to that of the folded equation by
\begin{eqnarray}
\frac{1}{\omega} \hat{F}_\omega^0 (Q_T, Q) & = & F_\omega^0 (Q_T,
Q) + \frac{\overline {\alpha}_S}{\omega} \int \frac{d^2 q}{\pi
q^2} \Theta (Q_T^2 - q^2) F_\omega^0 (Q_T, q) \nonumber \\
& & - \; \frac{\overline {\alpha}_S}{\omega} \Theta (Q_T^2 - Q^2)
\int_{Q^2}^{Q_T^2} \frac{d^2 q}{\pi q^2} q^2 \frac{\partial
F_\omega^0 (Q_T,q)}{\partial q^2} \log \left ( \frac{q^2}{Q_T^2}
\right ) \left
[ \left ( \frac{Q}{q} \right )^\omega - 1 \right ]
\label{a21}
\end{eqnarray}
\noindent The second term on the right hand side of (\ref{a20})
corresponds to real gluon emission without any \lq\lq
non-Sudakov" damping, but with angular ordering included (cf.
(\ref{a17})).  The following two terms are the unfolded virtual
corrections, which, if resummed, would lead to the non-Sudakov
form factor $\Delta_{NS}$.  The inhomogeneous term
$\hat{F}_\omega^0$ in the unfolded equation, (\ref{a20}), should
not contain any virtual corrections, but $F_\omega^0$ in the
folded equation, (\ref{a14}), may.

To solve (\ref{a21}) for $F_\omega^0$ in terms of
$\hat{F}_\omega^0$ it is simplest to use the following
representation for the driving term of the unfolded equation
(\ref{a20})
\begin{equation}
\frac{1}{\omega} \hat{F}_\omega^0 (Q_T, Q) = \int_0^1 dx \:
x^{\omega - 1} \int_x^1 \frac{dz}{z} \int \frac{d^2 q}{\pi q^2}
\Theta (Q - qz) \Phi^0 \biggl (\frac{x}{z}, |\mbox{\boldmath $q$}
+ \mbox{\boldmath $Q$}_T|, q \biggr ).
\label{b21}
\end{equation}
\noindent Then the solution of (\ref{a21}) is
\begin{equation}
F_\omega^0 (Q_T, Q) = \int_0^1 dx \: x^{\omega - 1} \int_x^1
\frac{dz}{z} \int \frac{d^2 q}{\pi q^2} \Theta (Q - qz)
\Delta_{NS} (z, q, Q_T) \Phi^0 \biggl (\frac{x}{z},
|\mbox{\boldmath $q$} + \mbox{\boldmath $Q$}_T|, q \biggr ).
\label{c21}
\end{equation}
\noindent The above representation, (\ref{b21}), mirrors relation
(\ref{a10}) which, for $n = 0$, expresses the driving term
$\hat{F}^0$ in terms of the non-perturbative input ${\cal F}_{n =
0}$.  In section 4 we will use (\ref{c21}) to specify the driving
term $F^0 (x, Q_T, Q)$ of the (folded) CCFM equation (\ref{a14})
in terms of an assumed form for the non-perturbative input
function $\Phi^0$.

\vspace*{3mm}
\noindent {\bf (b)  The BFKL limit}

We may take the leading $\log 1/x$ approximation of the unfolded
equation, (\ref{a20}), for the moments of the gluon distribution.
This corresponds to retaining the leading terms in the $\omega
\rightarrow 0$ limit.  That is we set $(Q/q)^\omega = 1$ in
(\ref{a20}), which then reduces to
\begin{equation}
F_\omega(Q_T, Q) = \frac{1}{\omega} \hat{F}_\omega^{0L} (Q_T, Q)
+ \frac{\overline{\alpha}_S}{\omega} \int \frac{d^2 q}{\pi q^2}
\left [F_\omega (|\mbox{\boldmath $Q$}_T + \mbox{\boldmath
$q$}|, q) - F_\omega (Q_T, q) \Theta (Q_T^2 - q^2) \right ]
\label{a22}
\end{equation}
\noindent where $\hat{F}_\omega^{0L}$ is the BFKL limit of
$\hat{F}_\omega^0$ of (\ref{a21}).  We see that the equation
(\ref{a22}) generates a moment function $F_\omega$ which is
independent of $Q$.  If we transform back from moment ($\omega$)
space to $x$ space, then we obtain the BFKL equation
\begin{equation}
\frac{\partial F(x, Q_T)}{\partial \log (1/x)} =
\overline{\alpha}_S \int \frac{d^2 q}{\pi q^2} \left [ F (x,
|\mbox{\boldmath $Q$}_T + \mbox{\boldmath $q$}|) - \Theta
(Q_T^2 - q^2) F (x, Q_T) \right ],
\label{a23}
\end{equation}
\noindent where we have neglected the derivative of the
inhomogeneous term with respect to $\log (1/x)$.  The two terms
in the integral on the right hand side correspond to real and
virtual gluon emission respectively.

\vspace*{3mm}
\noindent {\bf (c)  Cancellation of the real and virtual
singularities}

We first expose the cancellation of the singularities in the BFKL
limit \cite{jar}.  To do this we rewrite (\ref{a22}) with the
\lq\lq unresolved" real emissions (i.e. emissions with $q^2 <
\mu^2$) separated out
\begin{eqnarray}
F_\omega(Q_T) = \frac{1}{\omega} \hat{F}_\omega^{0L} (Q_T) & + &
\frac{\overline {\alpha}_S}{\omega} \int \frac{d^2 q}{\pi q^2}
\left [ F_\omega (|\mbox{\boldmath $Q$}_T + \mbox{\boldmath
$q$}|) \Theta (\mu^2 - q^2) - F_\omega (Q_T) \Theta (Q_T^2 -
q^2) \right ] \nonumber \\
& + & \frac{\overline {\alpha}_S}{\omega} \int \frac{d^2 q}{\pi
q^2} F_\omega (|\mbox{\boldmath $Q$}_T + \mbox{\boldmath $q$}|)
\Theta (q^2 - \mu^2).
\label{a24}
\end{eqnarray}
\noindent The singularities as $\mu^2 \rightarrow 0$ occur in the
second term on the right hand side.  To explicitly show the
cancellation between the real and virtual contributions, we
approximate the \lq\lq unresolved" real emission contribution
using $F_\omega(|\mbox{\boldmath $Q$}_T + \mbox{\boldmath $q$}|)
\simeq F_\omega(Q_T)$, which is valid at small $q$, and obtain
\begin{eqnarray}
\overline {\alpha}_S \int \frac{d^2q}{\pi q^2} \biggl[F_\omega
(|\mbox{\boldmath $Q$}_T + \mbox{\boldmath $q$}|) \Theta (\mu^2 -
q^2) - F_\omega (Q_T) \Theta (Q_T^2 - q^2) \biggr] \nonumber \\
= \; - \overline {\alpha}_S F_\omega (Q_T)
\int_{\mu^2}^{Q_T^2} \frac{dq^2}{q^2} + {\cal O}(\mu^2) \ \ \ \ \
\ \ \ \ \ \nonumber \\
\simeq \; - F_\omega(Q_T) \; \overline {\omega} (Q_T^2,\mu^2) \ \
\ \ \ \ \ \ \ \ \ \ \ \ \ \ \ \ \ \ \ \ \
\label{a25}
\end{eqnarray}
\noindent where
\begin{equation}
\overline {\omega}(Q_T^2, \mu^2) \equiv \overline {\alpha}_S
\log(Q_T^2/\mu^2).
\label{a26}
\end{equation}
\noindent The result (\ref{a25}) is the residual virtual
contribution to $\omega F_\omega(Q_T)$ which remains after the
cancellation of the real and virtual singularities.

If we substitute (\ref{a25}) into (\ref{a24}), we obtain
\begin{equation}
F_\omega(Q_T) = \; \frac{\hat{F}_\omega^{0L}(Q_T)}{\omega
+ \overline {\omega}} \; + \; \frac{\overline {\alpha}_S}{\omega
+ \overline {\omega}} \int \frac{d^2 q}{\pi q^2} \Theta (q^2 -
\mu^2) F_\omega(|\mbox{\boldmath $Q$}_T + \mbox{\boldmath $q$}|),
\label{a27}
\end{equation}
\noindent which, when we transform back to $x$ space, becomes
\begin{equation}
F(x, Q_T) = \int_x^1 \frac{dz}{z} z^{\overline \omega}
\hat{F}^{0L} \biggl (\frac{x}{z}, Q_T \biggr ) \; + \;
\overline{\alpha}_S \int_x^1 \frac{dz}{z} z^{\overline
\omega} \int \frac{d^2 q}{\pi q^2} \Theta (q^2 - \mu^2) F \biggl
(\frac{x}{z}, |\mbox{\boldmath $Q$}_T + \mbox{\boldmath $q$}|
\biggr ).
\label{b27}
\end{equation}
\noindent Equation (\ref{b27}) clearly corresponds to the \lq\lq
folded" BFKL equation in which we have resummed all the \lq\lq
unresolved" real emissions and {\bf all} the virtual corrections.

We recognise the Regge-like form $z^{\overline \omega}$ of the
non-Sudakov form factor $\Delta_{NS}^L$ in the folded BFKL
equation, which screens the $z \rightarrow 0$ singularities.  We
can also make the identification of the form factor using
(\ref{a8}), which in this case becomes
\begin{eqnarray}
\Delta_{NS}^L (z, Q_T, \mu^2) & = & \exp \biggr\{-
\overline{\alpha}_S \int_z^1 \frac{dz^\prime}{z^\prime} \int
\frac{dk^2}{k^2} \Theta (Q_T^2 - k^2) \Theta (k^2 - \mu^2)
\biggr\} \nonumber \\
& = & \exp \biggr\{ - \overline {\alpha}_S \log (1/z) \log
(Q_T^2/\mu^2) \biggr\} \nonumber \\
& = & z^{\overline {\omega}},
\label{a28}
\end{eqnarray}
\noindent where $\overline \omega$ is given in (\ref{a26}).  Note
that the driving term $\hat{F}_\omega^{0L}/\omega$ of the
unfolded BFKL equation (\ref{a24}) is free of $\mu^2 \rightarrow
0$ singularities, whereas the driving term of the folded version
of equation (\ref{a27}) or (\ref{b27}) contains, through
$\overline \omega$,
these $\mu^2$ singularities.

In the same way we can introduce a resolution $\mu^2$ into the
unfolded CCFM equation, (\ref{a20}), so that

\newpage

\begin{eqnarray}
F_\omega(Q_T,Q) & = & \frac{1}{\omega} \hat{F}_\omega^0(Q_T, Q) +
\nonumber \\
& & + \frac{\overline{\alpha}_S}{\omega} \int \frac{d^2 q}{\pi
q^2} \left [\Theta (Q - q) + \left (\frac{Q}{q} \right )^\omega
\Theta (q - Q) \right ] \Theta (q^2 - \mu^2) F_\omega
(|\mbox{\boldmath $Q$}_T + \mbox{\boldmath $q$}|, q) \nonumber \\
& & - \; \frac{\overline{\alpha}_S}{\omega} \int \frac{d^2 q}{\pi
q^2} \Theta (Q_T^2 - q^2) \Theta (q^2 - \mu^2) F_\omega (Q_T, q)
\nonumber \\
& & + \; \frac{\overline{\alpha}_S}{\omega} \Theta (Q_T^2 - Q^2)
\int_{Q^2}^{Q_T^2} \frac{d^2 q}{\pi q^2} q^2 \frac{\partial
F_\omega (Q_T,q)}{\partial q^2} \log \left (\frac{q^2}{Q_T^2}
\right ) \left [ \left ( \frac{Q}{q} \right )^\omega - 1 \right
].
\label{a29}
\end{eqnarray}
\noindent After resumming all the virtual corrections and \lq\lq
unresolved" radiation we obtain, in analogy with (\ref{a16})
\begin{equation}
F_\omega (Q_T,Q) = \tilde{F}_\omega^0 (Q_T, Q ; \mu^2) +
\overline {\alpha}_S \int \frac{d^2 q}{\pi q^2} \tilde{H}_\omega
(Q, Q_T, q ; \mu^2) F_\omega (|\mbox{\boldmath $Q$}_T +
\mbox{\boldmath $q$}|, q) \Theta (q^2 - \mu^2)
\label{a30}
\end{equation}
\noindent with
\begin{equation}
\tilde{H}_\omega (Q, Q_T, q ; \mu^2) = \int dz \; z^{\omega - 1}
\Theta (Q - qz) \tilde{\Delta}_{NS} (z, q, Q_T ; \mu^2)
\label{a31}
\end{equation}
\noindent where now the non-Sudakov form factor is of the form
\begin{equation}
\tilde{\Delta}_{NS} (z, q, Q_T ; \mu^2) = \exp \left ( -
\overline {\alpha}_S \int_z^1 \frac{dz^\prime}{z^\prime} \int
\frac{dk^2}{k^2} \Theta (Q_T^2 - k^2) \Theta (k - z^\prime q)
\Theta (k^2 - \mu^2) \right ).
\label{a32}
\end{equation}
\noindent The driving term of (\ref{a30}) has a similar form to
(\ref{c21}) except that the introduction of the resolution
cut-off leads to a factor $\Theta (q^2 - \mu^2)$ and to
$\Delta_{NS}$ being replaced by $\tilde{\Delta}_{NS}$ of
(\ref{a32})
\begin{equation}
\tilde{F}_\omega^0 (Q_T, Q; \mu^2) = \int_0^1 dx \; x^{\omega -
1} \int_x^1 \frac{dz}{z} \int \frac{d^2 q}{\pi q^2} \Theta (Q -
qz) \Theta (q^2 - \mu^2) \tilde{\Delta}_{NS} (z, q, Q_T; \mu^2)
\Phi^0 \biggl (\frac{x}{z}, |\mbox{\boldmath $q$} +
\mbox{\boldmath $Q$}_T|, q \biggr ).
\label{b32}
\end{equation}

In fact we do not need to introduce a lower cut-off $\mu^2$ on
the $dq^2$ integration in (\ref{a14}), since the equation remains
finite as $q \rightarrow 0$.  However, here we have demonstrated
the modifications necessary if, for pragmatic reasons, a cut-off
is introduced.  In particular we see $\Delta_{NS}$ of (\ref{a8})
must be replaced by $\tilde{\Delta}_{NS}$ of (\ref{a32}).  The
results will, of course, be independent of the choice of the
resolution $\mu^2$, up to ${\cal O}(\mu^2/Q_T^2)$.

Although the unfolded CCFM equation, (\ref{a20}), reduces to the
BFKL equation in the leading $\log (1/x)$ (or leading $\overline
{\alpha}_S/\omega$) approximation and although both equations are
free from singularities at $q^2 = 0$, the details of the
cancellations are different \cite{march3}.  In the BFKL case the
real emission terms and the virtual corrections are individually
divergent and have to be separately regulated by the $q^2 =
\mu^2$ cut-off.  On the other hand, for the CCFM equation both
terms are {\bf finite}, yet they generate additional powers of
$1/\omega$.  To be precise, when we solve eq. (\ref{a20})
iteratively we find after $n$ iterations that
\begin{equation}
F_\omega^{(n)} (Q_T, q) \sim \frac{1}{\omega^n} q^\omega \log^n
(q/Q_T)
\label{a33}
\end{equation}
\noindent as $q \rightarrow 0$.  The factor $q^\omega$ regulates
the integrals at $q^2 = 0$ and keeps both the real and virtual
terms finite, yet they separately contain double logarithmic
factors, $(\log^2 (1/x))^n$, arising from the behaviour of the
integral
\begin{equation}
\frac{1}{\omega^n} \int_0 \frac{dq^2}{q^2} q^\omega \log^n \left
(\frac{q^2}{Q_T^2} \right ) \sim \frac{1}{\omega^{2n + 1}}.
\label{a34}
\end{equation}
\noindent The $(\log (1/x))^{2n}$ behaviour exactly cancels
between the real emission and the virtual correction terms. \\

\vspace*{4mm}
\noindent {\large \bf 4.  Numerical solution of the CCFM
equation}

We explore the structure of the gluon distribution $F(x,Q_T,Q)$
at small $x$ by numerically solving the CCFM equation
(\ref{a14}).  To be precise we solve the equation in the presence
of the $\mu^2$ resolution cut-off (cf. (\ref{a30}))
\begin{equation}
F(x,Q_T,Q)  =  F^0 (x, Q_T, Q; \mu^2) + \int_x^1 \frac{dz}{z}
\int \frac{d^2 q}{q^2} \overline{\alpha}_S \Theta (Q - zq)
\Theta (q^2 - \mu^2) \tilde{\Delta}_{NS} (z,q,Q_T; \mu^2)
F(\frac{x}{z}, Q_T^\prime, q)
\label{a35}
\end{equation}
\noindent with $Q_T^\prime = |\mbox{\boldmath $Q$}_T + (1-z)
\mbox{\boldmath $q$}|$, and where the non-Sudakov form factor
$\tilde{\Delta}_{NS}$ is evaluated using (\ref{a32}).  We take
the argument of $\alpha_S$ to be $Q_T^2$ since this value is
usually assumed for small $x$ studies involving the BFKL
equation.

{}From (\ref{b32}) we see that the driving term is given by
\begin{equation}
F^0 (x, Q_T, Q; \mu^2) = \int_x^1 \frac{dz}{z} \int \frac{d^2
q}{\pi q^2} \Theta (Q - qz) \Theta (q^2 - \mu^2)
\tilde{\Delta}_{NS} (z, q, Q_T; \mu^2) \Phi^0 \biggl
(\frac{x}{z}, |\mbox{\boldmath $q$} + \mbox{\boldmath $Q$}_T|, q
\biggr ).
\label{a36}
\end{equation}

\noindent In general $\Phi^0$ will contain some smearing in $q$.
However, this is an inessential complication and it is sufficient
to assume strong-ordering in transverse momenta and to take the
$q$ dependence of $\Phi^0$ to be $\delta (\mbox{\boldmath $q$} +
\mbox{\boldmath $Q$}_T)$.  The $x/z$ and $Q_T$ dependence of
$\Phi^0$ are chosen so that if the $\Theta$ functions and
$\tilde{\Delta}_{NS}$ were to be set equal to 1 in (\ref{a36}),
then $F^0$ would reduce to $3(1 - x)^5 N \exp (-Q_T^2/Q_0^2)$.
The normalisation $N$ is fixed so that the gluon, integrated over
all $Q_T^2$, carries half the momentum of the proton.  $Q_0^2$ is
taken to be 1 GeV$^2$.  These assumptions are equivalent to the
choice
\begin{eqnarray}
F^0 (x, Q_T, Q; \mu^2) & = & N \exp (- Q_T^2/Q_0^2) \int_x^1
\frac{dz}{z} \Theta (Q - Q_T z) \Theta (Q_T^2 - \mu^2) \nonumber
\\
& & \tilde{\Delta}_{NS} (z, Q_T, Q_T; \mu^2) \frac{d [3 (1 -
x/z)^5 ]}{d \log (z/x)}.
\label{b36}
\end{eqnarray}

We solve (\ref{a35}) by iteration starting from the input form
given in (\ref{b36}).  We restrict the iterative procedure to the
domain\footnote{Strictly speaking the non-Sudakov form factor,
(\ref{a32}), is also dependent on $Q_0^2$ to ensure complete
cancellation of the real and virtual emissions.} $Q_T^2,
Q_T^{\prime 2} > Q_0^2 = 1 \: {\rm GeV}^2$.  We take the
upper limit cut-off, $Q_F^2$, on the $q^2$ integrations to be in
the region $10^4-10^5 \: {\rm GeV}^2$, though the results are
insensitive to variations around and above these values.  We use
a lower cut-off $\mu^2$, although we saw in Section 3 that the
CCFM equation is well behaved in the $\mu^2 \rightarrow 0$ limit.

Clearly, therefore, the results should be independent of the
choice of the resolution $\mu^2$, up to contributions of ${\cal
O}(\mu^2/Q_T^2)$ and ${\cal O}(\mu^2/Q^2)$.  This is well
demonstrated by the sample of results shown in Figs.\ 4 and 5
which correspond to the choices $\mu^2 = 10^{-2}, 10^{-1}, 0.5$
and, $1 \: {\rm GeV}^2$.  The convergence to a stable result with
decreasing $\mu^2$ is a test of the consistency of the solution.
{}From now on we show results for the choice $\mu^2 = 10^{-2} \:
{\rm GeV}^2$.

\vspace*{3mm}
\noindent {\bf (a)  CCFM solutions compared with those of the DLL
approximation}

Starting from the input of (\ref{b36}) we iterate (\ref{a35})
until a stable result for $F(x,Q_T,Q)$ is obtained.  The
numerical procedure is outlined in appendix B.  We take $Q_F^2 =
10^5 \: {\rm GeV}^2$.  The whole calculation is repeated using
the double-leading-logarithm (DLL) approximation in which we
replace angular ordering, $\Theta (Q - zq)$, by ordering in
transverse momenta, $\Theta (Q - q)$, and in which we set
$\Delta_{NS} = 1$.

Figs.\ 6, 7 and 8 respectively show a representative sample of
results for the $Q, Q_T$ and $x$ dependence of the gluon
distribution in the small $x$ regime.  The three plots of Fig. 6
compare the $Q^2$ dependence of $F(x, Q_T, Q)$ obtained from the
CCFM and DLL equations, for $Q_T^2 = 1, 10$ and 100 GeV$^2$
respectively.  We see that $F$ obtained from the CCFM equation is
less dependent on $Q^2$ than the DLL values of $F$.  In the DLL
case the suppression of $F$ in the region $Q^2 \lapproxeq Q_T^2$
is simply a reflection of the ordering of transverse momenta that
is embodied in the equation; an ordering which is not, in fact,
appropriate in the small $x$ domain.  Recall that in the BFKL
leading $\log (1/x)$ limit $F$ becomes independent of $Q^2$.  The
CCFM solutions in Fig.\ 6 exhibit this behaviour for the larger
values of $Q^2$, but for smaller $Q^2$ they show that non-leading
$\log (1/x)$ contributions begin to become important.  The low
$Q^2$ region is where the physically motivated angular ordering
embodied in the CCFM equation (but not in the BFKL equation)
provides more of a constraint.

Fig.\ 7 shows the $Q_T$ distributions of the gluon obtained from
solving the CCFM equation and from the approximate DLL equation.
As before the DLL solutions satisfy $Q_T^2 \lapproxeq Q^2$ even
at the smallest values of $x$, which again reflect the transverse
momentum ordering, $\Theta(Q - q)$, contained in equation
(\ref{b14}).  On the other hand the CCFM solutions become
significantly broader in $Q_T$, with decreasing $x$, on account
of the more appropriate angular ordering constraint $\Theta (Q -
zq)$.  The extensive $Q_T$ tail is a key property which
characterises the gluon distribution in the small $x$ domain.

Fig.\ 8 shows the behaviour of the integrated gluon distribution,
$xg (x,Q^2)$ of (\ref{a2}) with lower limit $Q_0^2$, as a
function of $x$ for various values of $Q^2$.  Note that the gluon
distributions are generated radiatively from an input which is
\lq\lq flat" at small $x$, (\ref{a36}), and so the rapid rise of
$xg$ with decreasing $x$ (shown as continuous curves) is
generated by the CCFM equation.  To quantify the increase, we
show in Fig.\ 9 the effective value of $\lambda$, defined by
\begin{equation}
xg (x, Q^2) = Ax^{-\lambda}.
\label{a37}
\end{equation}
\noindent For small $x$ we see that the solutions converge to a
typical $x^{-0.5}$ behaviour, approximately independent of $Q^2$,
which, as we shall see below, is consistent with that obtained
from the solution of the (leading $\log (1/x)$) BFKL equation.

The dashed curves in Figs.\ 8 and 9 show the characteristic
double-leading-logarithm (DLL) small $x$ behaviour
\begin{equation}
xg(x, Q^2) \sim \exp \biggl [ 2\{ \xi (Q^2, Q_0^2) \log (1/x)
\}^{\frac{1}{2}} \biggr ]
\label{t1}
\end{equation}
\noindent appropriate to a \lq\lq flat" input, that is $xg(x,
Q_0^2) \rightarrow$ constant as $x \rightarrow 0$.  In (\ref{t1})
we have omitted slowly varying functions of the argument of the
exponential.  We see that the steepness or \lq\lq effective
slope" $\lambda$ increases with $Q^2$ via the \lq\lq evolution
length"
\begin{equation}
\xi (Q^2, Q_0^2) = \int_{Q_0^2}^{Q^2} \frac{dq^2}{q^2}
\overline{\alpha}_S (q^2).
\label{t2}
\end{equation}
\noindent This behaviour is clearly evident in the DLL results of
Figs. 8 and 9.  The difference with the CCFM predictions for
small $x, x \lapproxeq 10^{-2}$, shows the importance of
implementing angular ordering in this domain.

\vspace*{3mm}
\noindent {\bf (b)  Comparison with solutions of the BFKL
equation}

To investigate the small $x$ limit we compare the CCFM solutions
with those that we obtained by solving the BFKL equation with the
same driving term.  To be precise we solve the unfolded BFKL
equation
\begin{eqnarray}
F (x, Q_T) & = & F^{0L} (x, Q_T) + \nonumber \\
& + & \overline{\alpha}_S (Q_T^2) \int_x^1 \frac{dz}{z}
\int_{Q_0^2}^\infty \frac{dQ_T^{\prime 2}}{Q_T^{\prime 2}} \left
\{ \frac{Q_T^{\prime 2} F (z, Q_T^\prime) - Q_T^2 F(z,
Q_T)}{|Q_T^{\prime 2} - Q_T^2|} + \frac{Q_T^2
F(z, Q_T)}{(4 Q_T^{\prime 4} +
Q_T^4)^{\frac{1}{2}}} \right \}
\label{a39}
\end{eqnarray}
\noindent with the driving term
\begin{equation}
F^{0L} (x, Q_T) = 3(1 - x)^5 N \exp (-Q_T^2/Q_0^2).
\label{a40}
\end{equation}
\noindent The integration region is restricted to $Q_T^{\prime 2}
> Q_0^2 = 1 \: {\rm GeV}^2$.  For completeness, we show in
Appendix C that, for fixed coupling $\alpha_S$ and $Q_0^2 = 0$,
the BFKL equation (\ref{a39}) can be rewritten in the form
\begin{equation}
F (x, Q_T) = F^{0L} (x, Q_T) + \overline{\alpha}_S \int_x^1
\frac{dz}{z} \int \frac{d^2 q}{\pi q^2} \biggl[ F(x,
|\mbox{\boldmath $Q$}_T + \mbox{\boldmath $q$}|) - \Theta (Q_T^2
- q^2) F(x, Q_T) \biggr ],
\label{a41}
\end{equation}
\noindent which we obtained previously, see (\ref{a23}).

Fig.\ 10 compares the properties of the integrated gluons
obtained from solving the CCFM and BFKL equations, while Fig.\ 9
shows the resulting effective slopes $\lambda$.  Since the
solution, $F(x, Q_T)$, of the BFKL equation is independent of
$Q$, the $Q^2$ dependence observed for $xg$ comes entirely from
the $Q_T$ integration of (\ref{a2}).  On the other hand the
solution $F(x, Q_T, Q)$ of the CCFM equation has an intrinsic $Q$
dependence arising from angular-ordering, $\Theta (Q - qz)$.  As
a consequence we see from Fig.\ 10 that the CCFM gluon evolves
faster with $Q^2$. From Fig.\ 9 we note that the effective
slopes $\lambda$ of the integrated CCFM and BFKL gluons are
remarkably similar at small $x$.  We conclude that the
next-to-leading $\log (1/x)$ effects included in the CCFM
formalism have a comparatively weak effect on the $x^{- \lambda}$
behaviour, although we note that the onset of the $x^{- \lambda}$
form is more delayed for the CCFM solution.

\vspace*{4mm}
\noindent {\large \bf 5.  Conclusions}

We have solved a unified equation for the unintegrated gluon
distribution which incorporates BFKL dynamics at small $x$ and
Altarelli-Parisi evolution at larger $x$.  We called it the CCFM
equation after its originators --- Catani, Ciafaloni, Fiorani and
Marchesini.  Starting from a driving term based on a \lq\lq flat"
$3(1 - x)^5$ gluon with a narrow $Q_T$ distribution, $\exp (-
Q_T^2/Q_0^2)$, we used an iterative procedure to find the $x,
Q_T$ and $Q$ dependence of the gluon.

We concentrated on the behaviour of the gluon in the small $x$
regime.  The key ingredients of the CCFM equation are the
angular-ordering of gluon emissions and the presence of a non-
Sudakov form factor.  We found that the CCFM equation
generates a gluon $F(x, Q_T, Q)$ with a singular $x^{-\lambda}$
behaviour, with $\lambda \simeq 0.5$, and a $Q_T$ distribution
which broadens and develops a significant tail as $x$ decreases.
Moreover the angular-ordering introduces a dependence of the
unintegrated gluon on the scale $Q$, especially at the lower
values of $Q^2$.  Sample results are shown in Figs.\ 6--9.  It is
convenient to display the $x$ dependence of the integrated
distribution, $xg$ of (\ref{a2}), although we should recall that
the physically relevant quantity at small $x$ is the unintegrated
distribution $F(x, Q_T, Q)$.

We compared the CCFM solutions with the conventional DLL
approximation in which angular-ordering is replaced by
strong-ordering in the gluon transverse momenta and in which the
non-Sudakov form factor is omitted, $\Delta_{NS} = 1$.  The gluon
is then found to be much less steep with decreasing $x$ and to
have a narrower $Q_T$ distribution.  We found that the DLL
approximation starts to differ from the CCFM results in the
region $x \lapproxeq 10^{-2}$.

We then compared the CCFM solutions with the solutions of the
BFKL approximation, based on an equation in which the
angular-ordering is ignored, and which therefore has unintegrated
solutions $F(x, Q_T)$ which do not depend on the scale $Q$.  In
fact the gluon $g(x, Q^2)$ obtained from the BFKL equation
acquires its $Q^2$ dependence entirely from the integration in
(\ref{a2}).  As a consequence we find that the integrated BFKL
solutions evolve more slowly in $Q^2$ than those obtained from
the CCFM equation, which have an additional intrinsic $Q^2$
dependence as shown, for example, in Fig.\ 6.  Fig.\ 9 quantifies
the $x^{-\lambda}$ agreement between the unified CCFM solution
and the approximate BFKL solution.  The agreement is remarkably
good at small $x$, especially at the larger values of $Q^2$.
Both the CCFM and BFKL solutions have a behaviour $xg \sim x^{-
\lambda}$ at small $x$, where the value of $\lambda$ is in the
region of 0.5 with only a modest dependence on $Q^2$, in contrast
to the dependence of $\lambda$ on the evolution length for the
DLL approximation, see Fig.\ 9.

It is appropriate to add a word of caution.  Our study has
concentrated on obtaining the perturbative QCD predictions of the
behaviour of the gluon at small $x$.  In particular we have made
the small $x$ approximation for the splitting function (\ref{a4})
\begin{equation}
\tilde{P} \simeq \overline {\alpha}_S \Delta_{NS} \frac{1}{z}
\label{a38}
\end{equation}
\noindent and we have set the Sudakov form factor $\Delta_S = 1$.

In other words we have neglected the singularity at $z = 1$ and
its Sudakov suppression.  Moreover we have neglected the effect
of the flavour singlet quarks on the evolution of the gluon.
Although these may be viewed as moderate-to-large $x$ effects,
they may, to some extent, feed through to small $x$ leading to
possible next-to-leading $\log (1/x)$ contributions to $F(x, Q_T,
Q)$.  Therefore, at this stage, our results should be regarded as
illustrative of the main properties of the gluon distribution at
small $x$, and as an indication of the respective domains of
validity of BFKL and conventional Altarelli-Parisi dynamics.  To
obtain a quantitative prediction of the gluon for all $x$ we
must include a proper treatment of the $z = 1$ behaviour and
include the quark distributions . \\

\vspace*{4mm}
\noindent {\large \bf Acknowledgements}

We thank G. Marchesini and B. R. Webber for valuable discussions.
J.\ K.\ thanks the Department of Physics and Grey College of the
University of Durham for their warm hospitality.  This work has
been supported in part by the UK Particle Physics and Astronomy
Research Council, the Polish KBN-British Council collaborative
research programme, the KBN grant no.\ 2 P302 062 04 and the EU
under contracts no.\ CHRX-CT92-0004/CT93-0357. \\

\vspace*{4mm}
\noindent{\large \bf Appendix A :  Unfolding the CCFM equation}

Here we derive the \lq\lq unfolded" version of the CCFM equation,
which is shown in (\ref{a20}).  We start by unfolding the
non-Sudakov form factor $\Delta_{NS}$ in (\ref{a19}).  The
relevant
term is
\begin{eqnarray}
\int_{Q^2}^\infty \frac{d^2 q}{\pi q^2} \left (\frac{Q}{q} \right
)^\omega \Delta_{NS} (z = Q/q, q, Q_T) F_\omega (|\mbox{\boldmath
$Q$}_T + \mbox{\boldmath $q$}|, q) \ \ \ \ \ \ \ \ \nonumber \\
= \: Q^\omega G_\omega (Q,Q_T) \: + \:
\int_{Q^2}^\infty \frac{d^2 q}{\pi q^2} \biggl(\frac{Q}{q}
\biggr)^\omega F_\omega (|\mbox{\boldmath $Q$}_T +
\mbox{\boldmath $q$}|, q)
\label{z1}
\end{eqnarray}
\noindent with
\begin{equation}
G_\omega (Q,Q_T) \equiv \int_{Q^2}^\infty \frac{d^2 q}{\pi q^2}
\biggl(\frac{1}{q^\omega} \biggr) [\Delta_{NS} (z = Q/q, q, Q_T)
- 1] F_\omega (|\mbox{\boldmath $Q$}_T + \mbox{\boldmath $q$}|,
q),
\label{z2}
\end{equation}
\noindent where we have simply subtracted and added 1 to
$\Delta_{NS}$.  For $Q_T^2 < Q^2$ the unfolding is trivial since,
then $\Delta_{NS}(z = Q/q, q, Q_T) = 1$, see (\ref{b8}).  This
leaves the case $Q_T^2 > Q^2$.  From the form of $\Delta_{NS}$ we
see that $G_\omega (Q_T, Q_T) = 0$, and so we may write
\begin{equation}
G_\omega (Q,Q_T) = \int_{Q^2}^{Q_T^2} \frac{dQ^{\prime
2}}{Q^{\prime 2}} \left [- \: \frac{\partial G_\omega (Q^\prime,
Q_T)}{\partial \log Q^{\prime 2}}
\right ].
\label{z3}
\end{equation}

\noindent If we differentiate (\ref{z2}) we obtain
\begin{eqnarray}
- \frac{\partial G_\omega (Q^\prime, Q_T)}{\partial \log
Q^{\prime 2}} & = & - \int_{Q^{\prime 2}}^\infty \frac{d^2 q}{\pi
q^2} \frac{1}{q^\omega} \frac{\partial \Delta_{NS} (z =
Q^\prime/q, q, Q_T)}{\partial \log Q^{\prime 2}} F_\omega
(|\mbox{\boldmath $Q$}_T + \mbox{\boldmath $q$}|, q) \nonumber \\
\nonumber \\
& = & \frac{1}{Q^{\prime \omega}} \int_{Q^{\prime 2}}^\infty
\frac{d^2 q}{\pi q^2} \biggl(\frac{Q^\prime}{q} \biggr)^\omega \:
\frac{\overline{\alpha}_S}{2} \log \biggl(\frac{Q^{\prime
2}}{Q_T^2} \biggr) \Delta_{NS} \biggl(z = \frac{Q^\prime}{q}, q,
Q_T \biggr) F_\omega (|\mbox{\boldmath $Q$}_T + \mbox{\boldmath
$q$}|, q) \nonumber \\ \nonumber \\
& = & \frac{1}{Q^{\prime \omega}} \log \biggl(\frac{Q^{\prime
2}}{Q_T^2} \biggr) \left [\frac{\partial F_\omega (Q_T,
Q^\prime)}{\partial \log Q^{\prime 2}} - \frac{\partial
F_\omega^0 (Q_T, Q^\prime)}{\partial \log Q^{\prime 2}} \right ]
\label{z4}
\end{eqnarray}
\noindent where the last equality is evident once we
differentiate (\ref{a16}).  Successively inserting (\ref{z4})
into (\ref{z3}), (\ref{z1}) and (\ref{a19}) we obtain the
$(Q/q)^\omega$ contribution to the last term in square brackets
in (\ref{a20}) and (\ref{a21}).

It remains to evaluate the term containing $F_\omega(Q_T, \min
\{Q,q\})$ in (\ref{a19}).  We may rewrite this term as
\begin{equation}
- \frac{\overline{\alpha}_S}{\omega} \int \frac{d^2 q}{\pi q^2}
\Theta (Q_T^2 - q^2) \biggl\{F_\omega (Q_T, q) \biggl[1 - \Theta
(q - Q) \biggr] + F_\omega (Q_T, Q) \Theta (q - Q) \biggr\}.
\label{z5}
\end{equation}
\noindent We immediately identify the term with 1 in the square
brackets as one of the terms in (\ref{a20}).  We are left with
two contributions from (\ref{z5}), each containing $\Theta (q -
Q)$, which combine to become
\begin{eqnarray}
\frac{\overline{\alpha}_S}{\omega} \int_{Q^2}^{Q_T^2} \frac{d^2
q}{\pi q^2} \biggl\{F_\omega (Q_T, q) - F_\omega (Q_T, Q)
\biggr\} = - \frac{\overline{\alpha}_S}{\omega}
\int_{Q^2}^{Q_T^2} \frac{d^2 q}{\pi q^2} \frac{\partial F_\omega
(Q_T, q)}{\partial \log q^2} \log \biggl(\frac{q^2}{Q_T^2}
\biggr),
\label{z6}
\end{eqnarray}
\noindent where the last equality follows on integrating by
parts.  We have now deduced (\ref{a20}), and (\ref{a21}), from
(\ref{a19}). \\

\vspace{4mm}
\noindent {\large \bf Appendix B :  Numerical technique used to
solve CCFM equation}

We briefly describe the numerical technique that we used to solve
the integral equation, (\ref{a35}), for the gluon distribution
$F(x, Q_T, Q)$.  The starting point is the (double) expansion of
the function $f (x, Q_T, Q) = Q_T^2 F(x, Q_T, Q)$ in terms of the
Tchebyshev polynomials with their arguments being linear
functions of the variables $\log (Q_T^2/\Lambda^2)$ and
$\log(Q^2/\Lambda^2)$.  To be precise we map the regions $Q_0^2 <
Q_T^2 < Q_F^2$ and $\mu^2 < Q^2 < Q_F^2$ into the interval (--1,
1) introducing respectively the variables
\begin{eqnarray}
\tau_T & = & 2 \log \left (\frac{Q_T^2}{Q_F Q_0} \right ) / \log
\left (\frac{Q_F^2}{Q_0^2} \right ) \nonumber \\ \nonumber \\
\tau & = & 2 \log \left (\frac{Q^2}{Q_F \mu} \right ) / \log
\left (\frac{Q_F^2}{\mu^2} \right ).
\label{y1}
\end{eqnarray}

\noindent We then expand the function $f$ in the polynomial form
\begin{equation}
f (x, Q_T, Q) = \sum_{i, j = 1}^N C_i(\tau_T) C_j(\tau) f_{ij}(x)
\label{y2}
\end{equation}
\noindent where the functions $f_{ij}(x)$ are the values of $f(x,
Q_T, Q)$ at the \lq\lq nodes" $Q_{Ti}^2$ and $Q_j^2$ specified by
\begin{equation}
\frac{Q_{Ti}^2}{Q_F Q_0} = \left (\frac{Q_F}{Q_0} \right
)^{\tau_i}, \hspace*{2cm} \frac{Q_j^2}{Q_F \mu} = \left
(\frac{Q_F}{\mu} \right )^{\tau_j},
\label{y3}
\end{equation}
\noindent with $\tau_k$ defined by
\begin{equation}
\tau_k = \cos \{ (k - {\textstyle \frac{1}{2}}) \pi/N \}.
\label{y4}
\end{equation}
\noindent The functions $C_k (\tau)$ in (\ref{y2}) are obtained
from the Tchebyshev polynomials $T_n (\tau)$ as follows
\begin{equation}
C_k (\tau) = \frac{1}{2 N} \sum_{n = 1}^N v_n T_n (\tau) T_n
(\tau_k)
\label{y6}
\end{equation}
\noindent where $v_1 = {\textstyle \frac{1}{2}}$ and $v_n = 1$
for $n > 1$.  In this way we can achieve a good approximation to
the $Q_T$ and $Q$ dependence of $f$ in terms of a modest number
$N$ of Tchebyshev polynomials.  Typically we take $N = 10$.

We now substitute the expansion (\ref{y2}) into the CCFM equation
(\ref{a35}) and obtain the following set of Volterra-type
integral equations for $f_{ij} (x)$
\begin{equation}
f_{ij} (x) = f_{ij}^0 (x) + \int_x^1 \frac{dz}{z} \sum_{k, \ell =
1}^N A_{ij, k\ell} (z) f_{k\ell} (x/z)
\label{y7}
\end{equation}
\noindent where the driving term
\begin{equation}
f_{ij}^0 (x) = Q_{Ti}^2 F^0 (x, Q_{Ti}, Q_j),
\label{y8}
\end{equation}
\noindent and the kernel
\begin{eqnarray}
A_{ij, k\ell} (z) & = & Q_{Ti}^2 \: \overline{\alpha}_S
(Q_{Ti}^2) \int_0^\pi d\phi \int \frac{dq^2}{\pi q^2
Q_{Ti}^{\prime 2}} \Theta (Q_j - qz) \Theta (Q_F^2 - q^2) \Theta
(q^2 - \mu^2)
\nonumber \\
& & \Theta (Q_{Ti}^{\prime 2} - Q_0^2) \tilde{\Delta}_{NS} (z, q,
Q_{Ti}, \mu^2) C_k (\tau) \ C_\ell (\tau).
\label{y9}
\end{eqnarray}
\noindent Here we have
\begin{equation}
Q_{Ti}^{\prime 2} = |\mbox{\boldmath $Q$}_T + \mbox{\boldmath
$q$} (1 - z)|^2 = Q_{Ti}^2 + (1 - z)^2 q^2 + 2(1 - z) qQ_{Ti}
\cos \phi
\label{y10}
\end{equation}
\noindent and the non-Sudakov form factor $\tilde{\Delta}_{NS}$
of (\ref{a32}).  The set of equations of (\ref{y7}) is solved
for the $f_{ij} (x)$ by iteration and the function $f (x, Q_T,
Q)$ then determined from (\ref{y2}). \\

\vspace*{4mm}
\noindent {\large \bf Appendix C :  Different forms of the BFKL
equation}

Here we show the equivalence of the forms of the BFKL equation
given in (\ref{a39}) and (\ref{a41}).  We start from (\ref{a39})
and write it in the form of a two-dimensional integration over
$d^2 Q_T^\prime$.  It is convenient to introduce a regulator
$\delta^2$ and to combine the two virtual contributions which
contain $F(z, Q_T)$.  Then (\ref{a39}), with $Q_0^2 = 0$, gives
\begin{eqnarray}
F(x, Q_T) & = & F^{0L} (x, Q_T) + \nonumber \\
& + & \lim_{\delta^2 \rightarrow 0} \overline{\alpha}_S \int_x^1
\frac{dz}{z} \int \frac{d^2 Q_T^\prime}{\pi}
\frac{1}{|\mbox{\boldmath $Q$}_T^\prime - \mbox{\boldmath
$Q$}_T|^2 + \delta^2} \left \{ F(z, Q_T^\prime) - \frac{Q_T^2
F(z, Q_T)}{[Q_T^{\prime 2} + |\mbox{\boldmath $Q$}_T^\prime -
\mbox{\boldmath $Q$}_T|^2 + \delta^2]} \right \}. \nonumber \\
& &
\label{x1}
\end{eqnarray}
\noindent We introduce an integration over the Feynman parameter
$\lambda$ to express the virtual contribution in (\ref{x1}) in
the form
\begin{eqnarray}
I & \equiv & Q_T^2 \int \frac{d^2 Q_T^\prime}{\pi}
\frac{1}{[|\mbox{\boldmath $Q$}_T^\prime - \mbox{\boldmath
$Q$}_T|^2 + \delta^2][Q_T^{\prime 2} + |\mbox{\boldmath
$Q$}_T^\prime - \mbox{\boldmath $Q$}_T|^2 + \delta^2]} \nonumber
\\
& = & Q_T^2 \int_o^1 d\lambda \int \frac{d^2 Q_T^\prime}{\pi}
\frac{1}{[\lambda Q_T^{\prime 2} + |\mbox{\boldmath $Q$}_T^\prime
- \mbox{\boldmath $Q$}_T|^2 + \delta^2]^2}.
\label{x2}
\end{eqnarray}
\noindent If we now make the substitution
$$
Q_T^\prime \rightarrow Q_T^\prime + \frac{Q_T}{\lambda + 1}
$$
\noindent then we can easily perform the $d^2 Q_T^\prime$ and
$d\lambda$ integrations and obtain
\begin{eqnarray}
I & = & \log \left ( \frac{Q_T^2}{\delta^2} \right ) + {\cal O}
(\delta^2/Q_T^2) \nonumber \\
& = & \int_{\delta^2}^{Q_T^2} \frac{d^2 q}{\pi q^2} + {\cal O}
(\delta^2/Q_T^2).
\label{x3}
\end{eqnarray}
\noindent We now substitute the virtual contribution $I$ of
(\ref{x3}) back into (\ref{x1}) and change the variable of
integration over the real emission contribution to
$\mbox{\boldmath $q$} = \mbox{\boldmath $Q$}_T^\prime -
\mbox{\boldmath $Q$}_T$.  In this way we obtain the form of the
BFKL equation shown in (\ref{a41}).

\newpage

\vspace*{4mm}
\noindent {\large \bf Figure Captions}
\begin{itemize}
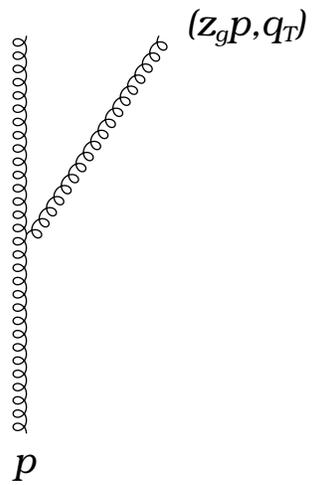
\item[Fig.\ 1] Gluon emission with probability given by
(\ref{a1}).
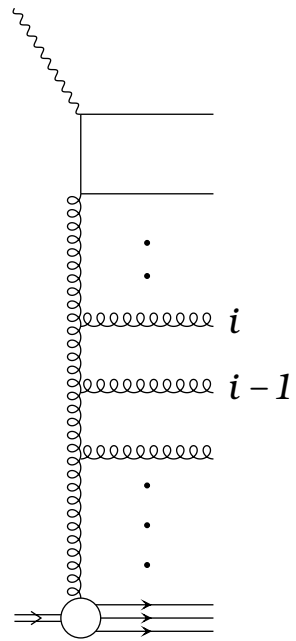
\item[Fig.\ 2] Multigluon emission in deep inelastic scattering
of a proton.
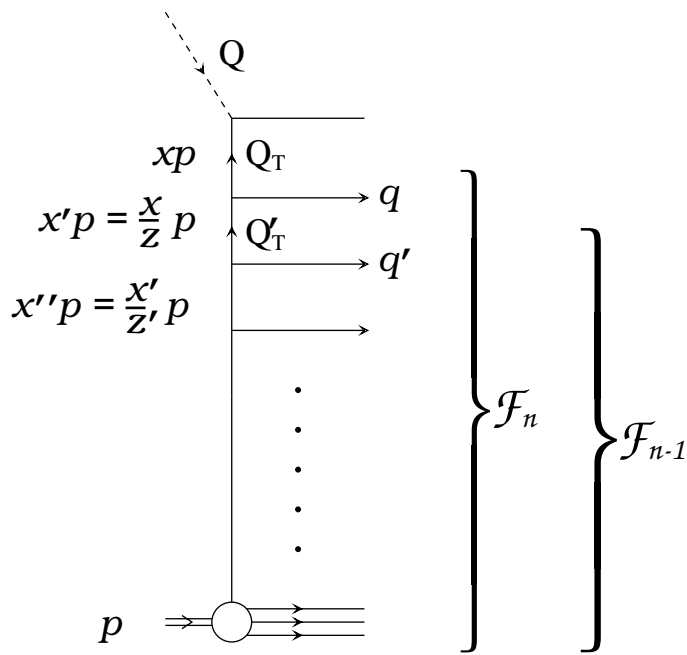
\item[Fig.\ 3] Kinematic variables for multigluon emission.  The
distribution ${\cal F}_n$ describing $n$ gluon emission is given
in terms of the distribution ${\cal F}_{n - 1}$ for $n - 1$ gluon
emission in (\ref{a6}).
\item[Fig.\ 4] The integrated gluon $xg(x, Q^2)$ calculated using
(\ref{a2}) from the solution $F(x, Q_T, Q)$ of the CCFM equation
(\ref{a35}) for different choices of the lower cut-off $q^2 =
\mu^2$ at each of five values of $Q^2$.  Here we take $Q_F^2 =
10^4 \: {\rm GeV}^2$.  Recall that our solutions are obtained
from a \lq\lq flat" gluon input.
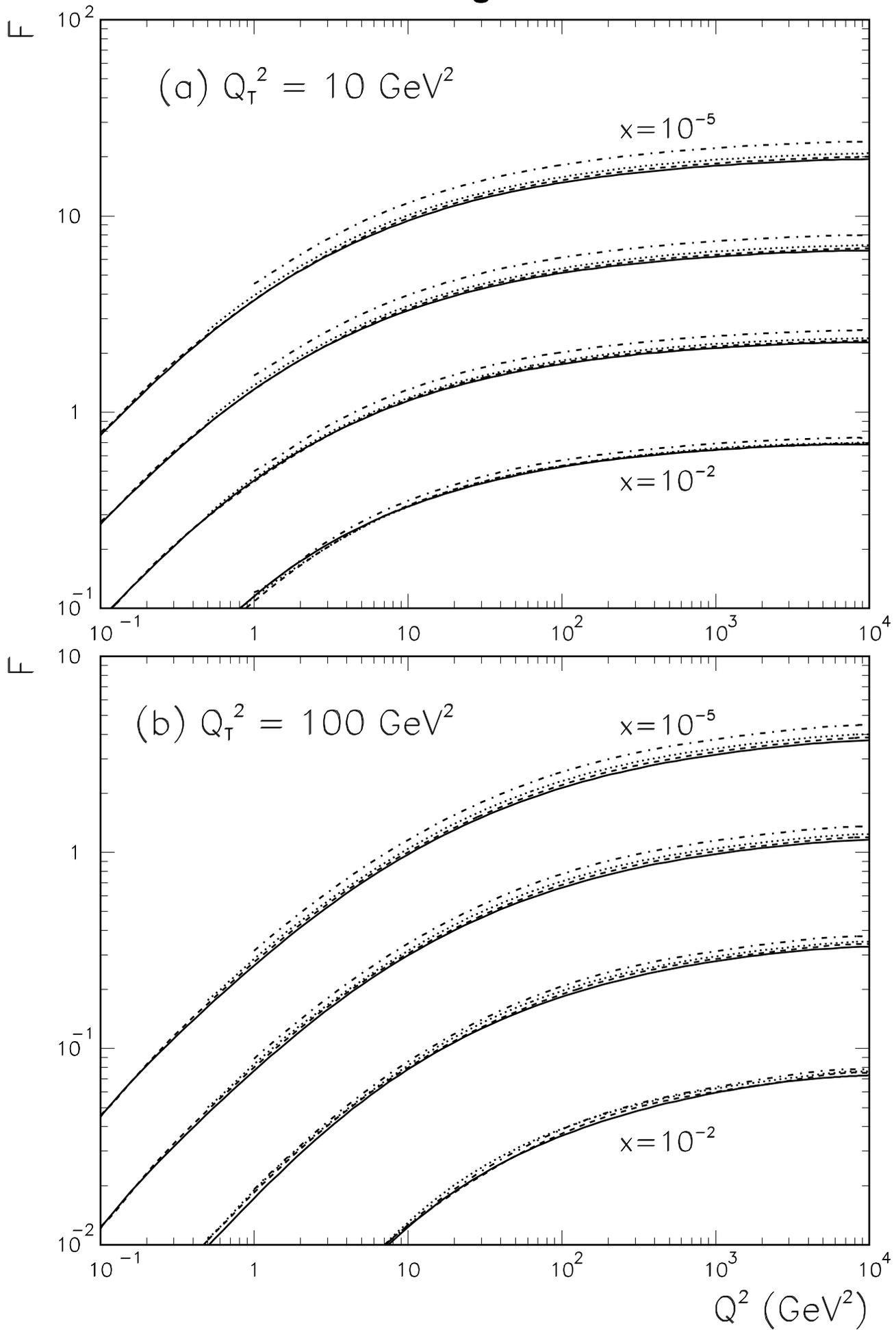
\item[Fig.\ 5] The $Q^2$ dependence of the gluon distribution
obtained by solving the CCFM equation (\ref{a35}) for different
choices of the lower cut-off $q^2 = \mu^2$ at each of two values
of $Q^2$ and four values of small $x$.  The values of
$\mu^2$ are as in Fig.\ 4.  Here we take $Q_F^2 = 10^4 \: {\rm
GeV}^2$.
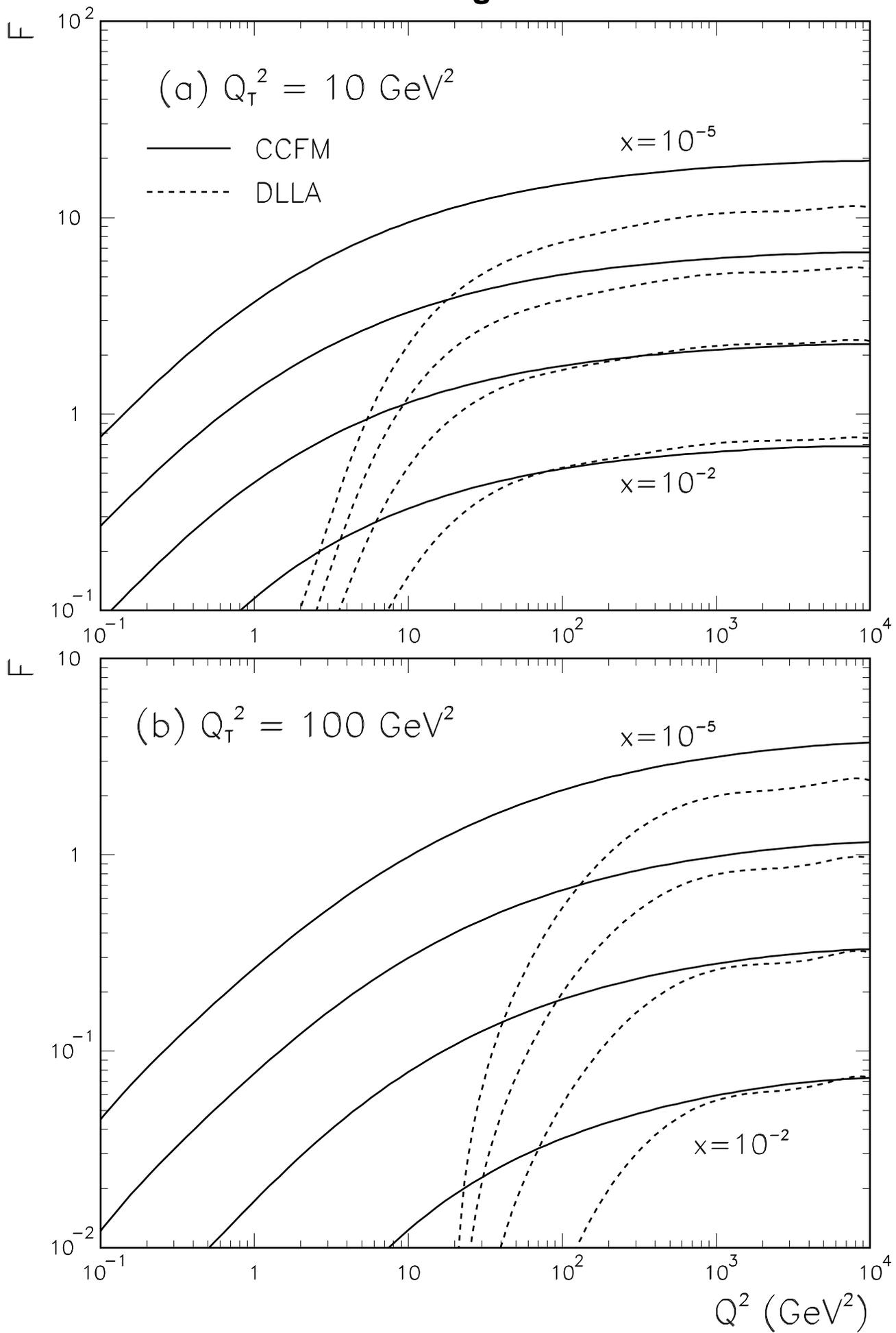
\item[Fig.\ 6] The $Q^2$ dependence of the gluon distribution
$F(x, Q_T, Q)$, obtained by solving the CCFM equation (\ref{a35})
(continuous curves), compared with that from the DLL equation
(dashed curves), for (a) $Q_T^2$ = 10 GeV$^2$ and (b) $Q_T$ = 100
GeV$^2$.  In each case we show the $Q^2$ dependence for $x =
10^{-5}, 10^{-4}, 10^{-3}$ and $10^{-2}$.
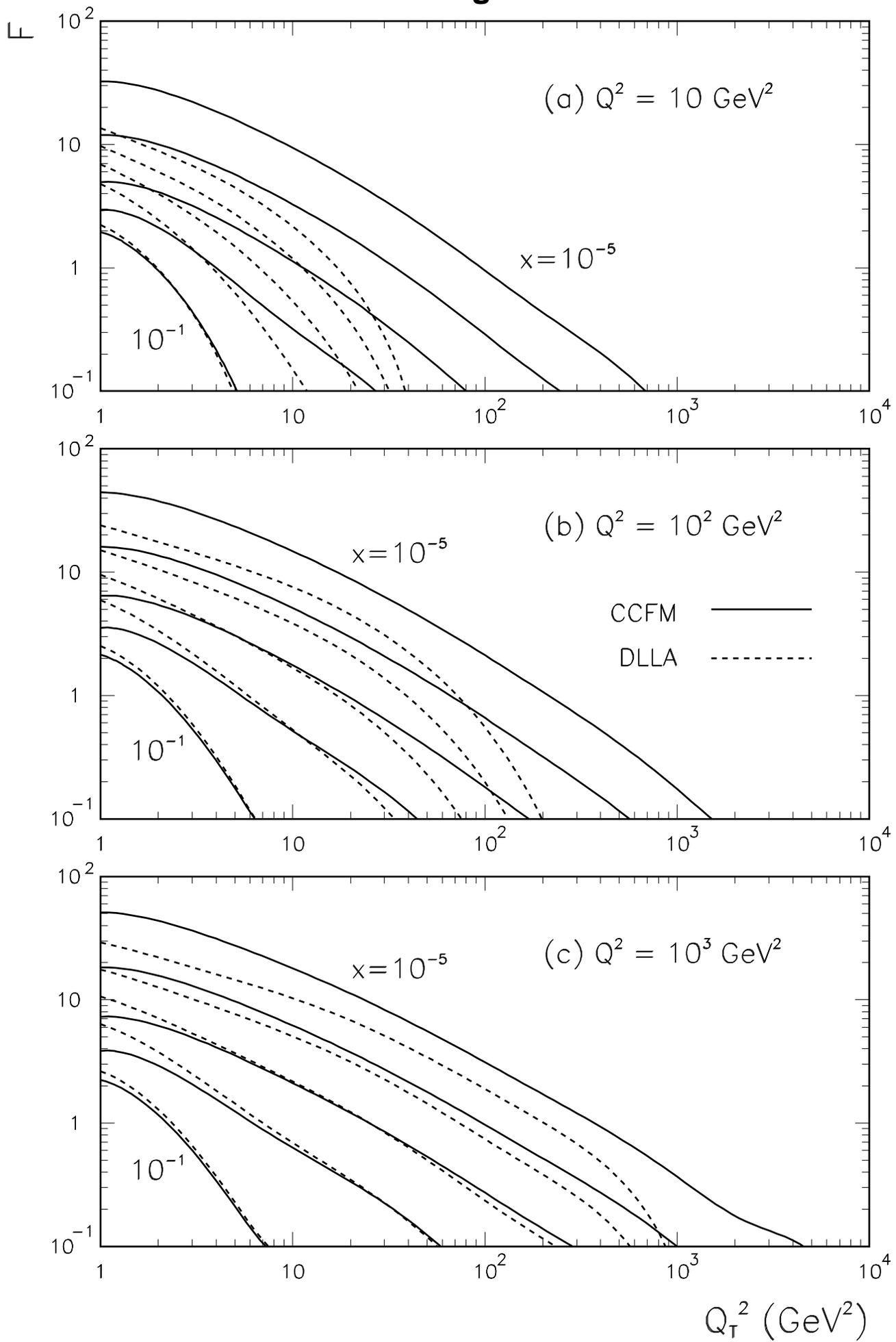
\item[Fig.\ 7] The $Q_T^2$ dependence of the gluon distribution
$F(x, Q_T, Q)$, obtained by solving the CCFM equation (\ref{a35})
(continuous curves) and the DLL equation (dashed curves), for (a)
$Q^2 = 10$ GeV$^2$, (b) $Q^2 = 10^2$ GeV$^2$ and (c) $Q^2 = 10^3$
GeV$^2$.  In each case we show curves for $x = 10^{-5}, 10^{-4},
10^{-3}, 10^{-2}$ and $10^{-1}$.
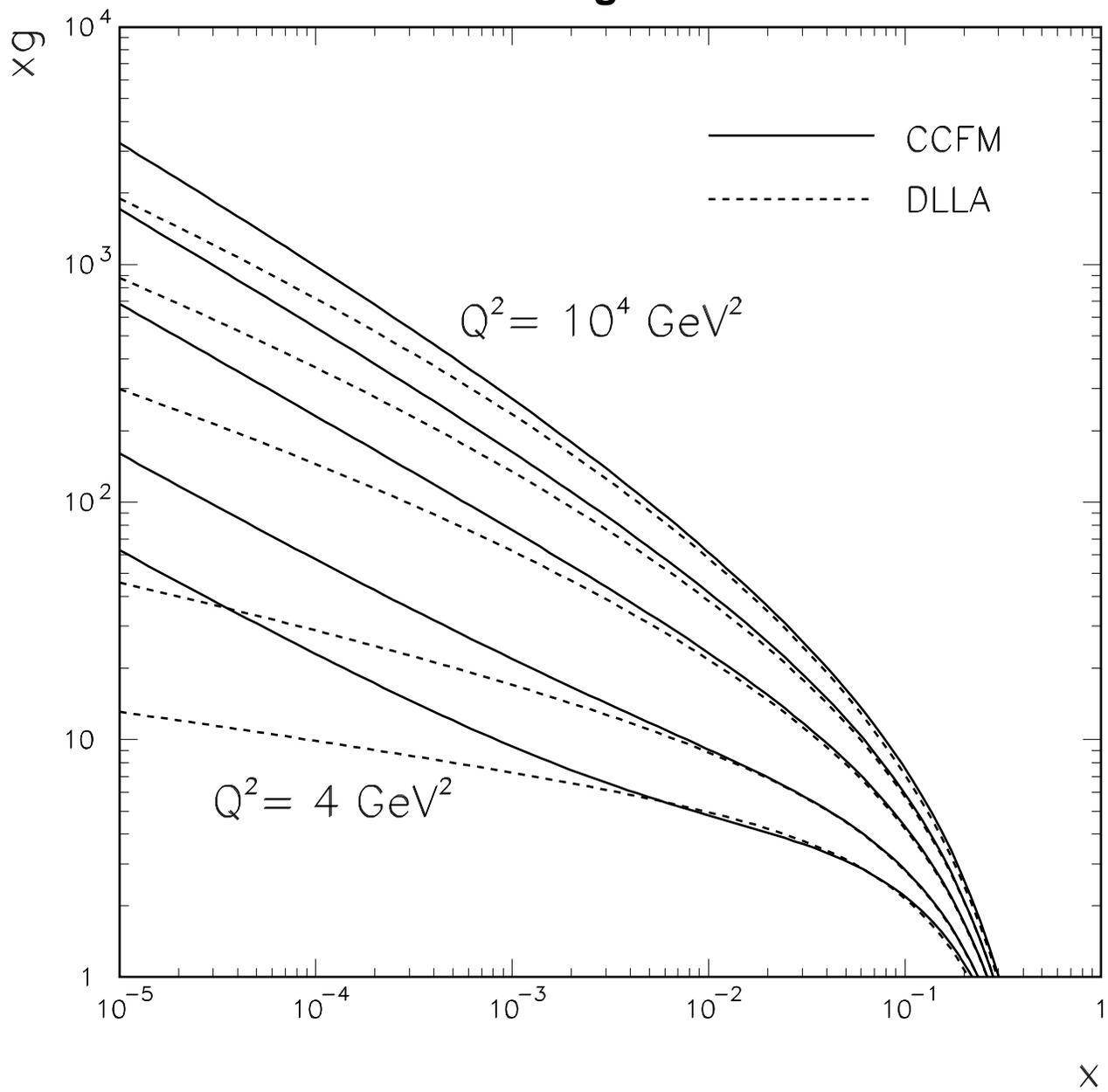
\item[Fig.\ 8] The integrated gluon distribution $xg$ versus $x$,
obtained from both the CCFM (continuous curves) and the DLL
(dashed curves) integral equations, for $Q^2 = 4, 10, 10^2, 10^3$
and $10^4$ GeV$^2$.  Recall that our solutions are obtained from
a \lq\lq flat" gluon input.
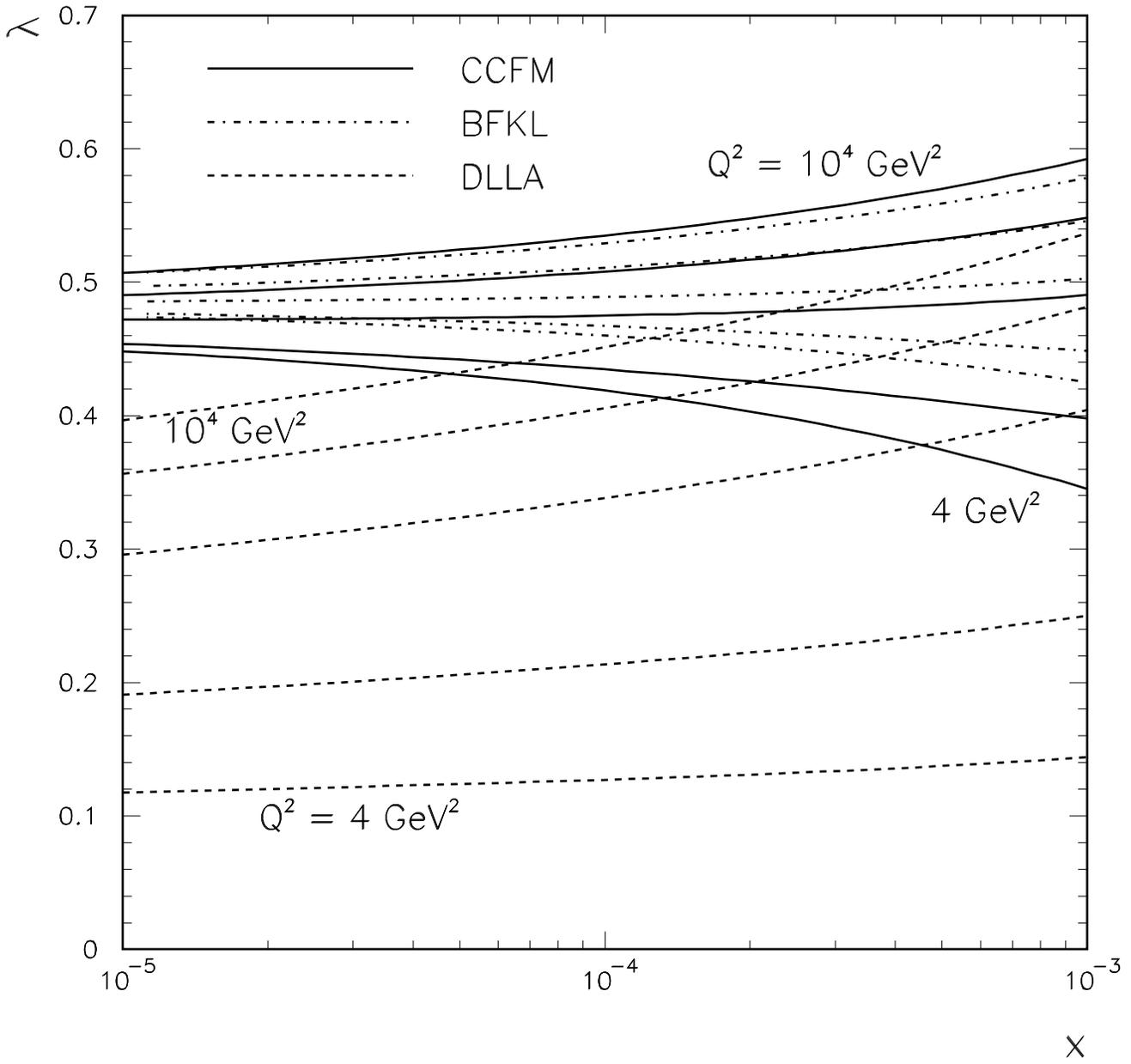
\item[Fig.\ 9] The effective values of $\lambda$, defined by $xg
= Ax^{-\lambda}$, obtained from the gluon distributions shown in
Figs.\ 8 and 10.  The CCFM values (continuous curves) are
compared with those obtained from the BFKL (dot-dashed curves)
and DLL approximations (dashed curves).  In each case we show
curves corresponding to the five different values of $Q^2$
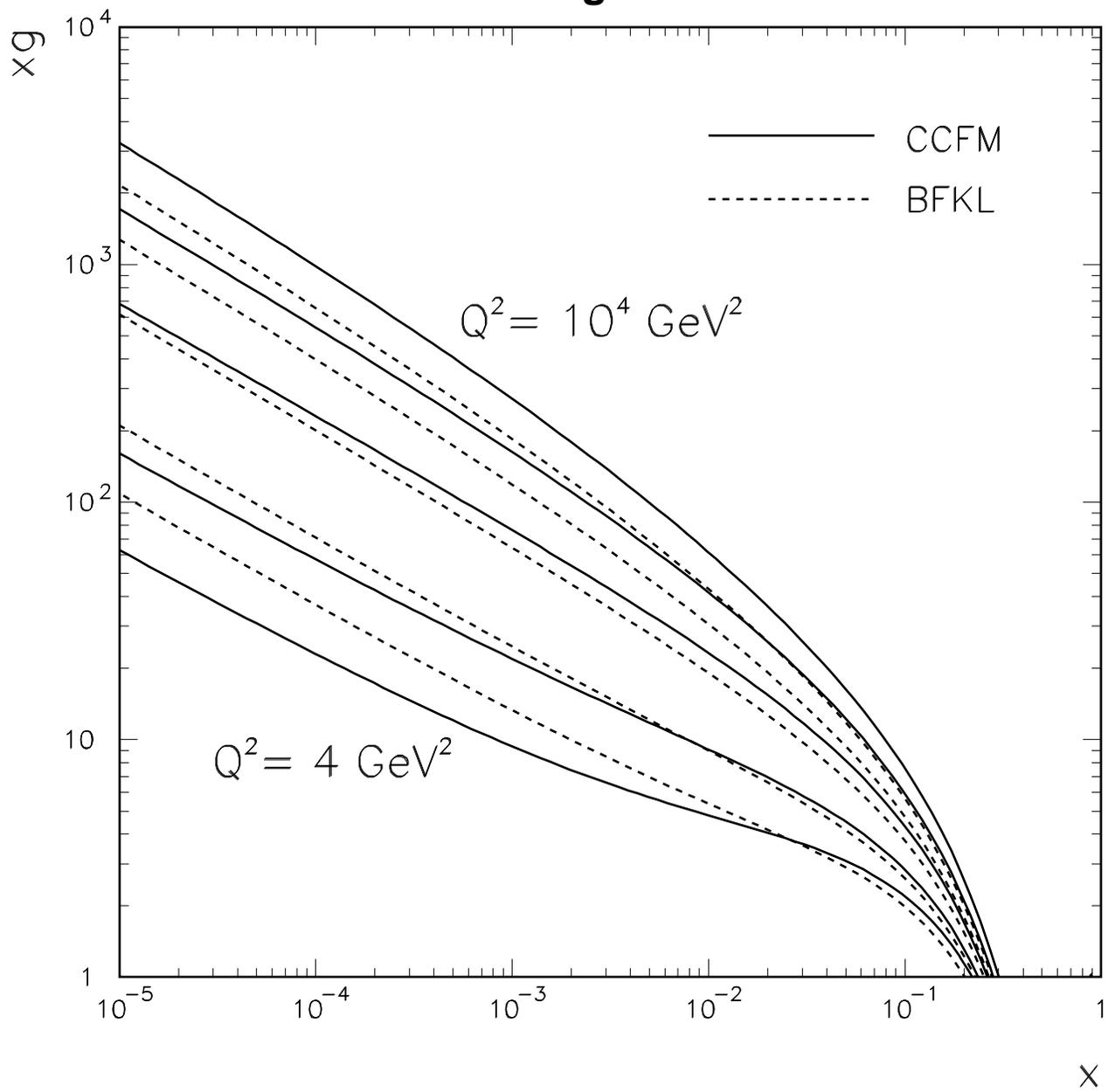
\item[Fig.\ 10] The integrated gluon distribution $xg$ versus
$x$, obtained from the CCFM (continuous curves) and the BFKL
(dashed curves) equations for $Q^2 = 4, 10, 10^2, 10^3$ and $10^4
\: {\rm GeV}^2$.  Recall that our solutions are obtained from a
\lq\lq flat" gluon input.

\end{itemize}

\end{document}